\documentclass{article}
\usepackage{emulateapj}
\usepackage{psfig}
\usepackage{times}

\def\lc{l_{\rm c}}
\def\lbb{l_{\rm bb}}
\def\tautr{\tau_{\rm tr}}
\def\tauc{\tau_{\rm c}}
\def\taue{\tau_{\rm e}}
\def\gta{\ga}
\def\lta{\la}
\def\Tbb{T_{\rm bb}}
\def\ledd{L_{\rm Edd}}

\def\msun{{\,M_\odot}}

\newcommand\fx{F_{\rm x}}

\newcommand\dm{\dot{m}}
\newcommand\fdisk{F_{\rm d}}
\newcommand\tcomp{T_{\rm comp}}

\newcommand\prad{P_{\rm rad}}

\newcommand\lnet{\Lambda_{\rm net}}

\newcommand\Tmin{T_{\rm min}}
\def\mean#1{\langle #1 \rangle}
\def\refindent{\par\noindent\hangindent=3pc\hangafter=1 }

\def\aasup#1#2#3{\refindent#1, A\&AS, #2, #3}

\def\apj#1#2#3{\refindent#1, {\it ApJ}, {\bf#2}, #3.}
\def\apjlett#1#2#3{\refindent#1, {\it ApJL}, {\bf #2}, #3.}
\def\apjsup#1#2#3{\refindent#1, ApJS, #2, #3}

\def\mnras#1#2#3{\refindent#1, {\it MNRAS}, {\bf#2}, #3.}
\def\nature#1#2#3{\refindent#1, {\it Nature}, {\bf #2}, #3.}

\def\refpaper#1#2#3#4{\refindent#1, #2, #3, #4}
\def\refbook#1{\refindent#1}

\def\>{$>$}
\def\<{$<$}

\def\simlt{\lower.5ex\hbox{$\; \buildrel < \over \sim \;$}}
\def\simgt{\lower.5ex\hbox{$\; \buildrel > \over \sim \;$}}
\def\sqr#1#2{{\vcenter{\hrule height.#2pt
      \hbox{\vrule width.#2pt height#1pt \kern#1pt
         \vrule width.#2pt}
      \hrule height.#2pt}}}

\begin{document}

\centerline{Submitted to the Astrophysical Journal}

\title{X-rays From Magnetic Flares In Cygnus X-1: The Role Of A Transition
  Layer}

\author{Sergei Nayakshin$^1$ and  James B. Dove$^2$}

\affil{$^1$Physics Department, The University of Arizona,
Tucson AZ 85721\\ 
$^2$CASA, University of Colorado, Boulder, CO 80309}

\begin{abstract}
  The spectrum of Seyfert 1 Galaxies is very similar to that of
  several Galactic Black Hole Candidates (GBHCs) in their hard state,
  suggestive that both classes of objects have similar physical
  processes. While it appears that an accretion disk corona (ADC)
  model, where the corona sandwiches the cold accretion disk, is
  capable of explaining the observations of Seyfert galaxies, recent
  work has shown that this model is problematic for GBHCs. To address
  the differences in the spectra of Seyferts and GBHCs, we consider
  the structure of the atmosphere of the accretion disk in a region
  near an active magnetic flare (we refer to this region as transition
  layer). We show that the transition layer is subject to a thermal
  instability, similar in nature to the ionization instability of
  quasar emission line regions.  We find that due to the much higher
  ionizing X-ray flux in GBHCs, the only stable solution for the upper
  layer of the accretion disk is that in which it is highly ionized
  and is at the Compton temperature ($kT \sim $ a few keV).  Using
  numerical simulations for a slab geometry ADC, we show that the
  presence of a transition layer, here modeled as being completely
  ionized, can significantly alter the spectrum of escaping radiation
  for modest optical depths.  Due to the higher albedo of the disk,
  the transition layer leads to a reduction in reprocessing features,
  i.e., the iron line and the X-ray reflection hump.  In addition, if
  most of the accretion energy is dissipated in the corona, the
  thermal blackbody component is reduced, giving rise to a lower
  Compton cooling rate within the corona.  Although a transition layer
  does allow for higher coronal temperatures, global two-phase,
  slab-geometry ADC models still cannot have coronal temperatures high
  enough to explain the data unless the optical depth of the
  transition layer is unphysically large ($\gta 10$).  However, a
  model having a patchy corona rather than a global corona appears
  promising. Thus, it is possible that due to the thermal
  instability of the surface of the accretion disk, which leads to
  different endpoints for GBHCs and Seyfert galaxies, the X-ray
  spectra from these two types of objects can be explained by a single
  unifying ADC model.
\end{abstract}


\section{Introduction}
The X-ray spectra of Seyfert Galaxies and Galactic Black Hole
Candidates (GBHCs) indicate that the reflection and reprocessing of
incident X-rays into lower frequency radiation is an ubiquitous and
important process. For Seyfert Galaxies, the X-ray spectral index
hovers near a ``canonical value'' ($\sim 0.95$; Pounds et al. 1990,
Nandra \& Pounds 1994; Zdziarski et al. 1996, but also see Brandt \&
Boller 1998), after the reflection component has been subtracted out
of the observed spectrum. It is generally believed that the
universality of this X-ray spectral index may be attributed to the
fact that the reprocessing of X-rays within the cold accretion disk
leads to an electron cooling rate that is roughly proportional to the
heating rate inside the active regions (AR) (Haardt \& Maraschi 1991,
1993; Haardt, Maraschi \& Ghisellini 1994; Svensson 1996).

Although the X-ray spectra of GBHCs are similar to that of Seyfert
galaxies, they are considerably harder (most have a power-law index of
$\Gamma \sim 0.7$), and the reprocessing features are less prominent
(Zdziarski et al. 1996, Dove et al. 1997). It is the relatively hard
power law (and therefore the required large coronal temperature) and
the weak reprocessing/reflection features that led Dove et al.  (1997,
1998), Gierlinski et al. (1997) and Poutanen, Krolik \& Ryde (1997) to
conclude that the two-phase accretion disk corona (ADC) model, in both
patchy and slab corona geometry cases, does not apply to Cygnus X-1.

One of the main problems with the global slab-geometry ADC model is
that, for a given coronal optical depth, no self-consistent coronal
temperature is high enough to produce a spectrum both as hard and with
an exponential cutoff at an energy as high as that of Cyg~X-1 (Dove,
Wilms, \& Begelman 1997). However, this result is sensitive to the
assumption that the accretion disk is relatively cold, such that $\sim
90$\% of the reprocessed coronal radiation is re-emitted by the disk
as thermal radiation (with a temperature $\sim 150$ eV). It is this
thermal radiation that dominates the Compton cooling rate within the
corona.  If the upper layers of the accretion disk were highly
ionized, creating a ``transition layer,'' a smaller fraction of the
incident coronal radiation would be reprocessed into thermal radiation
(i.e., the albedo of the disk would be increased), and therefore the
Compton cooling rate in the corona would be reduced.  Furthermore, as
shown by Ross, Fabian \& Brandt (1996; RFB96 hereafter), the high
ionization state of the outer atmosphere of the accretion disk in
GBHCs may explain the weakness of observed iron line features in Cyg
X-1.

Recently, Nayakshin \& Melia (1997) investigated (via simple qualitative
considerations) the structure of the X-ray reflecting material in AGNs
assuming that the ARs are magnetic flares above the disk (Haardt et
al. 1994; see also Galeev, Rosner \& Vaiana 1979).  They showed that the
pressure and energy equilibrium conditions for the X-ray illuminated upper
layer of the disk require the gas temperature to be $T\sim$ few $\times
10^5$ K, leading the upper atmosphere of the disk to a low ionization
state.

In this paper, we consider the reprocessing of X-rays from ARs for
GBHCs. In \S2, we show that there should be a thermal instability at
the surface of the cold disk, causing the temperature to climb up to
$T\sim$ a few $\times 10^7$ K. This temperature is roughly the Compton
temperature with respect to the coronal radiation field.  At this
temperature the transition layer turns out to be almost completely
ionized. Recently, B\"ottcher, Liang, and Smith (1998), using an
iterative method, a linear Comptonization algorithm, and the
photoionization model XSTAR, also found that a highly ionized
transition layer should form for moderate ionization parameters,
i.e., ionization parameters thought to be appropriate for GBHCs.
In \S3, we explore the ramifications of this highly ionized transition
layer on the energetics of the corona, and investigate how it alters
the spectrum of escaping radiation. We also discuss whether slab
geometry ADC models, when transition layers are included, can account
for the observed spectra of BHCs. In \S4, we give our conclusions.

\section{The Formation of a Transition Layer}

\subsection{Physical Conditions in X-ray Skin Near Magnetic Flares}
\label{sect:pressure}

\begin{figure*}[t]
\centerline{\psfig{file=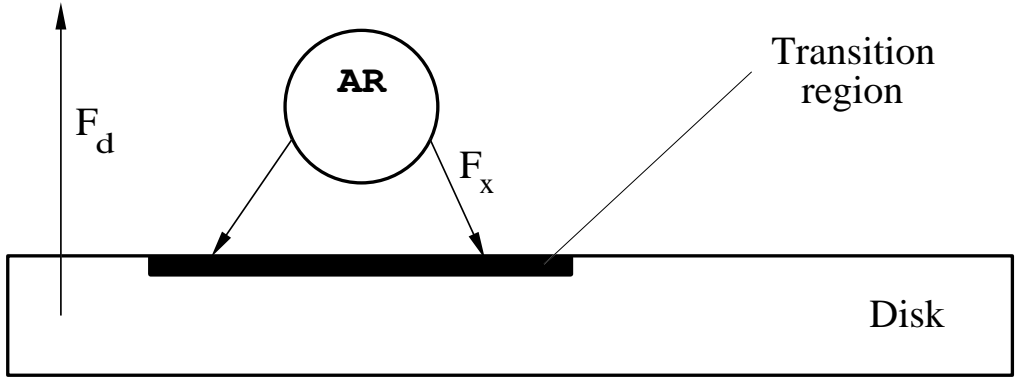,width=.72\textwidth}}
\caption{The geometry of the active region (AR: magnetic flare) and
the transition layer. Magnetic fields, containing AR and supplying it
with energy are not shown. Transition region is defined as the  upper layer
of the disk with Thomson depth of $\sim$ few, where the incident X-ray flux
is substantially larger than the intrinsic disk flux}
\label{fig:geometry}
\end{figure*}

The two-phase model was put forward by Haardt \& Maraschi (1991, 1993)
to explain the spectra of Seyfert Galaxies. Haardt et al. (1994)
pointed out that observations are inconsistent with a uniform corona
and introduced a patchy corona, where each ``patch'' is a magnetic
flare (also referred as an active region). The key assumptions of the
model are (1) during the flare, the X-ray flux from the active region
greatly exceeds the disk intrinsic flux, and (2) the compactness
parameter $l$ of the active region is large, so that the dominant
radiation mechanism is Comptonization of the disk thermal
radiation. We wish to apply the same model to the X-ray spectra of
GBHCs, and we employ the same assumptions.

As discussed below, we argue for the existence of a transition layer
in the vicinity of an active coronal region (see Figure 1). Since the
flux of ionizing radiation is proportional to $1/d^2 \times \cos i
\propto d^{-3}$, where $i$ is the angle between the normal of the disk
and the direction of the incident radiation field and $d$ is the
distance between the active region and the position on the disk, the
ionization state of the disk surface will vary across the disk.
Consequently, only the gas near the active regions (with a radial size
$\sim$ a few times the size of the active region, situated directly
below it) may be highly ionized. Most reprocessing of coronal
radiation will take place in the transition regions; in addition, most
radiation emitted by the disk that propagates through the active
regions will have been emitted in the vicinity of the transition
regions. Therefore, in this paper, we will only consider a one-zone
model, consisting of the active region, the transition layer, and the
underlying cold disk.

The structure of the transition layer, i.e., its temperature, density and
ionization state are determined by solving the energy, ionization and
pressure balance conditions. The first two conditions have been extensively
treated in the literature, and we will follow these standard methods here
(see \S \ref{sect:instability}). The pressure equilibrium condition is
typically replaced by the constant gas density assumption (e.g., RFB, Zycki
et al. 1994 and references therein). For the problem at hand, however, the
equilibrium state of the transition layer is sensitive to the pressure
balance, and thus we will attempt to take it into account. Accounting for
the pressure balance leads to results that are substantially different from
previous work, as discussed below.

The X-ray radiation pressure on the transition layer is equal to $\fx/c$,
where $\fx$ is the flux produced by the magnetic flare. The compactness
parameter of the active region, $l$, is defined as
\begin{equation}
l\equiv {\fx\sigma_T \Delta R\over m_e c^3},
\label{compact}
\end{equation}
and is expected to be larger than unity (e.g., Poutanen \& Svensson
1996, Poutanen, Svensson \& Stern 1997). The size of the active region
$\Delta R$ is thought to be of the order of the accretion disk height
scale $H$ (e.g., Galeev et al.  1979). Here, $H/R$ is estimated from
the gas pressure dominated solution of Svensson \& Zdziarski (1994;
SZ94 hereafter),
\begin{equation}
\frac{H}{R} = 7.5\times 10^{-3} (\alpha M_1)^{-1/10} r^{1/20}
 [\dot{m}J(r)]^{1/5}[(1-f)]^{1/10},
\end{equation}
where $\alpha$ is the viscosity parameter, $M_1\equiv M/10\msun$ is
the mass of the black hole, $f$ is the fraction of accretion power
dissipated into the corona (averaged over the whole disk), $r$ is the
radius relative to the Schwarzschild radius ($r\equiv R/R_g$), $J(r) =
1-(3/r)^{1/2}$.  For the case of the hard state of Cyg X-1, most of
the bolometric luminosity is in the hard X-ray band (e.g., Gierlinski
et al. 1997). Thus, most of the accretion energy must be dissipated
directly in the corona, i.e., $f\sim 1$ (Haardt and Maraschi 1991;
Stern et al. 1995). The dimensionless accretion rate $\dm =
\eta\dot{M}c^2/\ledd$ for Cyg~X-1 seems to be around $0.05$. Here,
$\dot{M}$ is the accretion rate, $\eta =0.056$ is the efficiency for
the standard Shakura-Sunyaev disk, and $\ledd$ is the Eddington
luminosity. Note that this definition of $\dm$ is different by factor
$\eta$ from that used by SZ94 (i.e., $\dm \simeq 17 \times \dm_{\rm
SZ94}$). Finally, for $r = 6$, the X-ray flux is
\begin{equation}
\fx \,\simeq \, 4 \times 10^{23} \, l\,\alpha^{1/10} M_1^{-9/10}
\left({\dm\over 0.05}\right)^{-1/5} (1-f)^{-1/10} {\rm erg \
cm}^{-2}\,{\rm sec}^{-1}.
\label{xflux}
\end{equation}

To check whether assumption (1) of the patchy two-phase model is consistent
for Cyg~X-1 parameters, we estimate the intrinsic flux of the disk,
\begin{equation}
\fdisk = 1.0 \times 10^{22} M_1^{-1} \left({\dm\over 0.05}\right)
(1-f)\;{\rm erg \ cm}^{-2}\,{\rm sec}^{-1}.
\label{diskflux}
\end{equation}
It is seen that the X-ray flux is indeed much larger than the
intrinsic disk emission if $1-f\ll 1$ and the compactness parameter
$l\gg 0.01$. 

The pressure of the disk surface layer before the occurrence of a
flare (or, equivalently, far enough from the flare), assuming that the
upper layer of the disk with Thomson optical depth $\tau_x\sim $ few
is in the hydrostatic equilibrium, is
\begin{eqnarray}
  P_0 &\simeq& {G M m_p\over R^2} \tau_x\, {H\over R} = 6.2\times 10^{10}\;
  M_1^{-11/10} \alpha^{-1/10}\nonumber \\ & &\times \tau_x\,\left({\dm\over
      0.05}\right)^{1/5} (1-f)^{1/10} \;{\rm erg \ cm}^{-3},
\label{p0}
\end{eqnarray}
where $r=6$ (SZ94). Near an active magnetic flare, the
ratio of the incident radiation pressure to the unperturbed accretion
disk atmosphere pressure is
\begin{equation}
{\fx\over c P_0}\, = \, 2. \times 10^2 \; l \,\tau_x^{-1}\,\left(\alpha
M_1\right)^{1/5} \,\left({\dm\over 0.05}\right)^{-2/5}
(1-f)^{-1/5},
\label{prat}
\end{equation}
Since the radiation pressure from the active region greatly exceeds the
unperturbed thermal pressure, the transition layer will contract until
the gas pressure $P$ 
\begin{equation}
P \sim \fx/c \; .
\label{pestm}
\end{equation}
A better treatment would solve for the gas opacity in the transition layer
and thus the radiation force self-consistently, rather than simply using the
ram pressure $\fx/c$ (Nayakshin 1998). However, inequality (\ref{pestm})
turns out to be sufficient to prove the main point of this paper. We thus
move on to solve the energy and ionization balance equations for the
transition layer.

\subsection{The Thermal Instability}\label{sect:instability}

A general condition for a thermal instability was discovered by Field
(1965). He argued that a physical system is usually in pressure
equilibrium with its surroundings. Thus, any perturbations of the
temperature $T$ and the density $n$ of the system should occur at a
constant pressure.  The system is unstable when
\begin{equation}
\left({\partial \Lambda_{\rm net}\over \partial T}\right)_{P} < 0,
\label{field}
\end{equation}
where the ``cooling function,'' $\lnet$, is the
difference between cooling and heating rates per unit volume, divided by
the gas density $n$ squared.

In ionization balance studies, it turns out convenient to define two
parameters. The first one is the ``density ionization parameter'' $\xi$, 
equal to (Krolik, McKee \& Tarter 1981)
\begin{equation}
\xi = {4\pi \fx\over n},
\label{xid}
\end{equation}
The second one is the  ``pressure ionization parameter'', defined as
\begin{equation}
\Xi \equiv \frac{\prad}{P},
\label{xip}
\end{equation}
where $P$ is the full isotropic pressure due to neutral atoms, ions,
electrons and the trapped line radiation. (This definition of $\Xi$ is
the one used in the ionization code XSTAR, and is more appropriate for
studies of the instabilities than the original definition of $\Xi$
given in Krolik et al. 1981, who used $n k T$ instead of the full
pressure $P$. For temperatures typical of GBHCs, however, we found
that the trapped line radiation was always a small fraction, e.g.,
$\sim$ few percent of the total pressure, and therefore the two
definitions of $\Xi$ are nearly identical).  Krolik et al. (1981)
showed that the instability criterion (\ref{field}) is equivalent to
\begin{equation}
\left( {d\Xi\over d T}\right)_{\lnet=0}\, < 0  \; ,
\label{fcond}
\end{equation}
where the derivative is taken with the condition $\lnet = 0$
satisfied, i.e., when the energy balance is imposed. In this form, the
instability can be easily seen when one plots temperature $T$ versus
$\Xi$.

\begin{figure*}
\centerline{\psfig{file=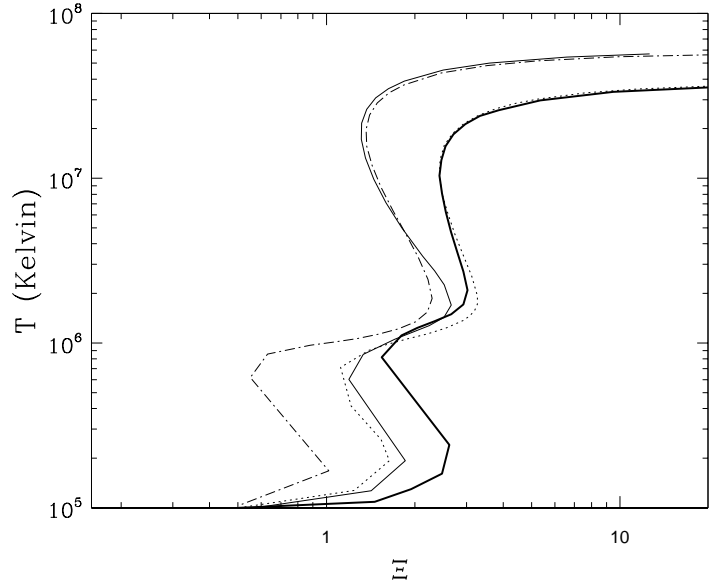,width=.5\textwidth,angle=0}}
\caption{Gas temperature versus pressure ionization parameter $\Xi$ --
  the ionization equilibrium curves for parameters appropriate for
  GBHCs.  The incident spectrum is approximated by a power law of
  photon index $\Gamma$, exponentially cutoff at $100$ keV, and the
  reflected blackbody with equal flux and temperature $\Tmin$
  (equation \ref{tmincyg}). Values of the parameters are: $\Gamma =$
  1.5, 1.75, 1.75, 1.7 and $k \Tmin = $ 200, 100, 200, 400 eV,
  corresponding to the fine solid, thick solid, dotted and dash-dotted
  curves, respectively. Because the ionization equilibrium is unstable
  when the curve has a negative slope, and because there exists no
  solution below $\Tmin$ (see text), the only stable solution is the
  uppermost branch of the curve with $T\simgt 10^7$ K.}
\label{fig:scurve}
\end{figure*}

We now apply the X-ray ionization code XSTAR (Kallman \& McCray 1982,
Kallman \& Krolik 1986), to the problem of the transition layer. A
truly self-consistent treatment would involve solving radiation
transfer in the optically thick transition layer, and, in addition,
finding the distribution of the gas density in the transition layer
that would satisfy pressure balance. Since radiation force acting on
the gas depends on the opacity of the gas, this is a difficult
non-linear problem. Thus, we defer such a detailed study to future
work, and simply solve (using XSTAR) the local energy and ionization
balance for {\em an optically thin layer} of gas in the transition
region. We assume that the ionizing spectrum consists of the incident
X-ray power law with the energy spectral index typical of GBHCs in the
hard state, i.e., $\Gamma = 1.5-1.75$, exponentially cutoff at 100
keV, and the blackbody spectrum from the cold disk below the
transition layer. If the energy and ionization balance is found to be
unstable for this setup, the transition layer will also be unstable,
because the instability is local in character.

Note that it is not possible for the transition region to have a
temperature lower than the effective temperature of the X-radiation, i.e.,
$\Tmin = (\fx/\sigma)^{1/4}$. The reason why the simulations may give
temperatures lower than $\Tmin$ for low values of $\xi$ is that in this
parameter range XSTAR neglects certain non-radiative de-excitation
processes, which leads to an overestimate of the cooling rate (Zycki et
al. 1994; see their section 2.3). Since we are using a one-zone model for
the transition layer, we use an attenuated X-ray flux $\mean{\fx}$ which
represents the surface-averaged value as seen by the transition region,
$\mean{\fx} = \fx/b = 0.1 \fx/b_1$, where $b$ is a dimensionless number of
order 10, $b_1 = b/10$, and $\fx$ is the X-ray flux leaving the active
region (see figure \ref{fig:geometry}). Using equation (\ref{xflux}),
\begin{equation}
\Tmin\simeq 5.0 \times 10^6\; l^{1/4} \, b_1^{-1/4}\, \left({\dm\over
0.05}\right)^{-1/20}\, \alpha^{1/40} \, M_1^{-9/40} \,
\left[1-f\right]^{-1/40}
\label{tmincyg}
\end{equation}

Figure \ref{fig:scurve} shows results of our calculations for several
different X-ray ionizing spectra. A stable solution for the transition
layer structure will have a positive slope, and also satisfy the
pressure equilibrium condition. As discussed in \S
\ref{sect:pressure}, $P \leq \fx/c$ (i.e., $\Xi\geq 1$). In addition,
if the gas is completely ionized, the absorption opacity is negligible
compared to the Thomson opacity. Because all the incident X-ray flux
is eventually reflected, the net flux is zero, and so the net
radiation force is zero (note that the momentum of the incident
radiation is reflected deep inside the disk, so the radiation does
apply a ram pressure equal to $2\fx/c$ to the whole disk, but {\em not
to the transition layer}). In that case the pressure $P$ adjusts to
the value appropriate for the accretion disk atmosphere in the absence
of the ionizing flux (see also Sincell \& Krolik 1996), which is given
by equation (\ref{p0}). Therefore, the pressure ionization parameter
should be in the range
\begin{equation}
1< \Xi < 1\times 10^2 \,l \left(\alpha
    M_1\right)^{1/5} \,\left({\dm\over 0.05}\right)^{2/5} (1-f)^{-1/5}.
\label{xip}
\end{equation}
With respect to the ionization equilibria shown in Figure
(\ref{fig:scurve}), the gas is almost completely ionized on the upper
stable branch of the solution (i.e., the one with $T\simgt 10^7$ K),
and thus the pressure equilibrium for such temperatures requires $\Xi
\sim \fx/c P_0\gg 1$, which is allowed according to equation
(\ref{xip}).

In addition to the stable Compton equilibrium state, there is a small
range in parameter space, with the temperature between $100$ and $200$
eV, in which a stable thermal state exists . The presence of this
region is explained by a rapid decrease in {\em heating} (with $T$
increasing), rather than an increase in cooling (cf. equation
\ref{field}). The X-ray heating rate decreases in the temperature
range $100-200$ eV with increasing $T$ since higher temperatures lead
to higher ionization rates of ion species with ionization energies
$\sim kT$. This larger degree of ionization reduces the X-ray opacity,
and thus the heating rate as well. Note that it is highly unlikely that
the transition region will stabilize within the temperature range
$100$ -- $200$ eV because the effective temperature, $\Tmin$, is most
likely above this temperature range.

\begin{figure*}
\centerline{\psfig{file=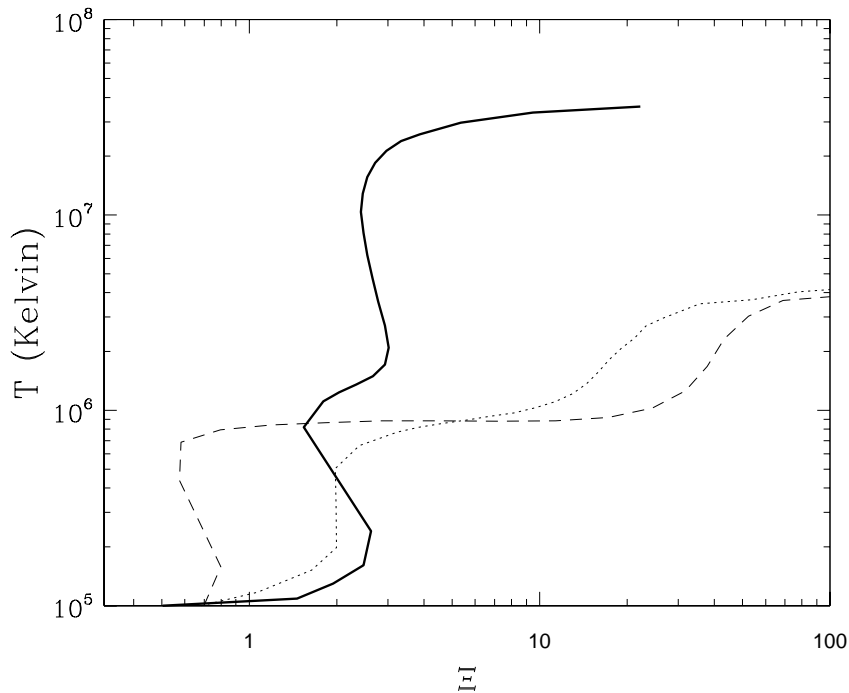,width=.6\textwidth,angle=0}}
\caption{Same as Figure \ref{fig:scurve}. The thick solid curve is
same as that in Figure \ref{fig:scurve} and is appropriate for the
hard state of a GBHC, whereas two other curves are relevant to the
soft state in GBHCs, or at large depth in the transition layer (see
Discussion). Values of the parameters are: $\Gamma =$ 2.1, 2.1 and $k
\Tmin = $ 200, 400 eV for the dotted and dashed curves, respectively.}
\label{fig:ssoft}
\end{figure*}

Due to the above considerations, it is very likely that the transition
layer is highly ionized in GBHCs {\em in the hard state} for $\tau_{\rm x}
\lta 1$. The upper limit of $\tau_{\rm x}$ can only be found by a more
exact treatment.  In addition, the transition layer may be heated by the
same process that heats the corona above it, albeit with a smaller heating
rate.  Furthermore, Maciolek-Niedzwiecki, Krolik \& Zdziarski (1997) have
recently shown that the thermal conduction of energy from the corona to the
disk below may become important for low coronal compactness parameters and
substantially contribute to the heating rate of the transition layer. Thus,
the transition layer may be even hotter than found by photoionization
calculations.

Eventually, the X-rays are down-scattered and the radiation spectrum becomes
softer as one descends from the top of the transition layer to its bottom.
We can qualitatively test the gas ionization stability properties by
allowing the ionizing spectrum to be softer than the observed spectrum of
GBHCs in the hard state. In Figure \ref{fig:ssoft} we show two examples
of such calculations. The slope of the ionization equilibrium curve becomes
positive everywhere above $k T\sim$ 100 eV, so that these equilibria are
stable, and thus the gas temperature may saturate at $T\sim \Tmin$, far
below few keV, the appropriate temperature for the uppermost layer of the
transition region.  Thus, we know (see also \S \ref{sec:discussion}) that the
transition layer should terminate at some value of $\tau_x \sim$ few.

We point out the similarity of our results to those of Krolik, McKee
\& Tarter (1981) for the emission line regions in quasars. These
authors solved the ionization and energy balance equations for
optically thin clouds illuminated by a broad band quasar spectrum, and
showed that the thermal instability exists if the pressure ionization
parameter is close to unity.  They found that there are two stable
states for the line emitting clouds, one cold and one hot. The cold
state corresponds to the gas temperature $T\lesssim $ few $\times
10^4$ K, where collisional cooling balances ionization heating. The
hot state corresponds to temperatures $T\gtrsim 10^7$ K, when
ionization heating decreases due to almost complete ionization of the
elements and the Compton heating decreases since the gas is close to
the Compton temperature $\tcomp$. Here, we found a similar instability
for the X-ray skin near an active magnetic flare in accretion disks of
GBHCs. The lower temperature equilibrium state is not allowed,
however, since the gas density and the X-ray ionizing flux in GBHCs is
larger than these quantities in quasar emission line regions by some
$\sim 10 -12$ orders of magnitude. Finally, we note that the thermal
instability is not apparent in studies where the gas density is fixed
to a constant value. Following Field (1965), we argue that the
assumption of the constant gas density is not justified for real
physical systems, and that one always should use the pressure
equilibrium arguments to determine the actual gas density and the
stability properties of the system.

\section{Global ADC Models with a Transition Layer} 
\subsection{Physical Setup}
\begin{figure*}
\centerline{\psfig{file=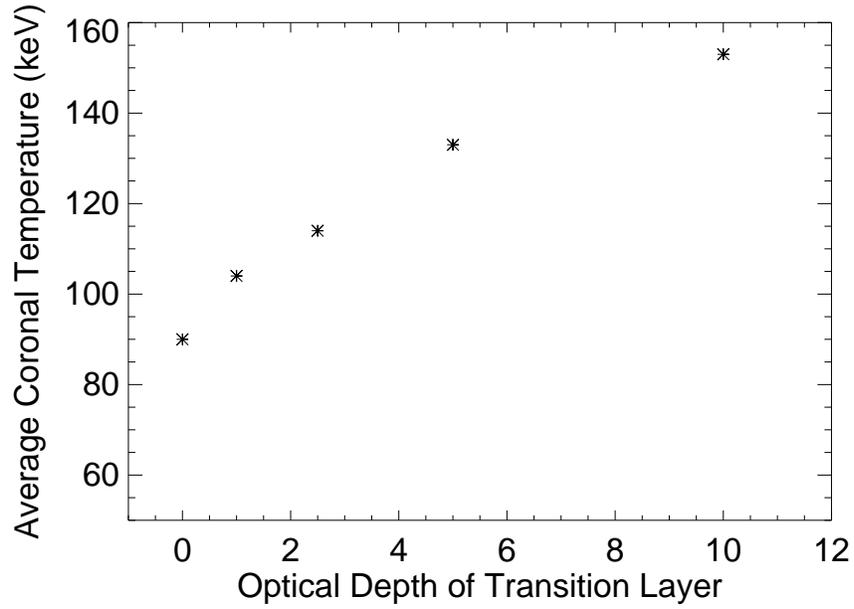,width=.6\textwidth,angle=0}}
\caption{The maximum temperature of the corona as a function of the optical
  depth of the transition layer. For all models, we assume that the
  intrinsic compactness parameter of the disk is $\lbb = 0.01$ and the disk
  temperature is $k\Tbb = 150$ eV. For all models, the maximum temperature
  is reached for $\lc \sim 2-5$.}
\label{fig:Tmax-vs-tautrans}
\end{figure*}

We now explore how a transition layer affects the physical properties of
the coronal gas and the spectrum of escaping radiation.  Since a transition
layer may occur for both a global ADC model and for a patchy coronal model,
both models should be explored in detail. In this paper, however, we only
study the global ADC model, and defer a self-consistent treatment of the
patchy corona to a future paper.  We expect, however, that the systematic
trends found for the global ADC models should also occur for the patchy
model, with the main difference being the maximum coronal temperature of
the patchy model being higher.

A self consistent treatment of the transition layer would solve for the
ionization layer structure as a function of optical depth.  However, we
will make the simplifying assumption that the transition layer is
completely ionized. The optical depth of the transition layer will be kept
as a free parameter, and we will explore how sensitive the spectrum of
escaping radiation is to $\tautr$. The two issues we will address are (1)
for a given coronal optical depth (e.g., one thought to be appropriate for
fitting the spectra of GBHCs), how does the maximum coronal temperature
depend on the optical depth of the transition layer, and (2) how does the
spectrum of escaping radiation, most importantly the reprocessing features,
vary with the optical depth of the transition layer.  Of particular
importance is the parameter range of $\tautr$ such that the predicted
spectrum is consistent with that of GBHCs, and whether these values of
$\tautr$ are consistent with the assumption of the transition layer being
completely ionized.
\begin{figure*}
\centerline{\psfig{file=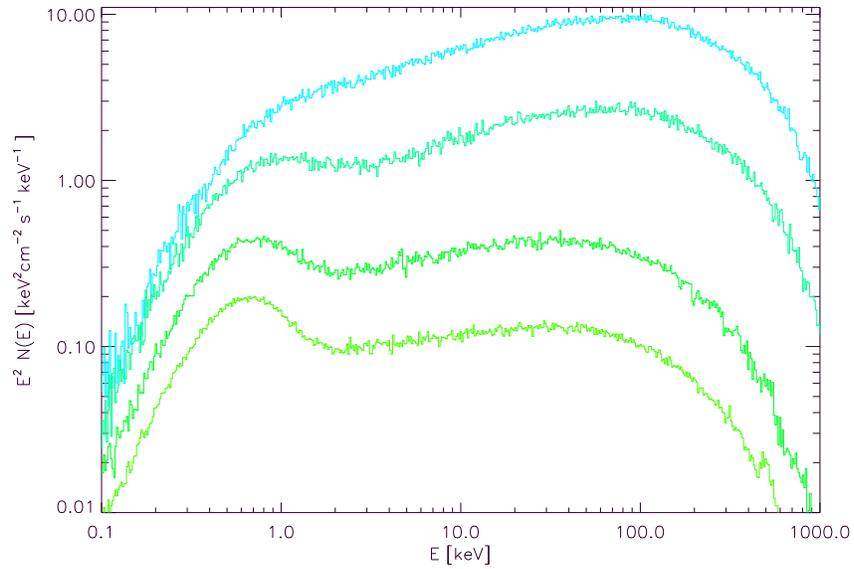,width=.6\textwidth}}
\caption{The predicted spectrum for various values of
the transition layer optical depth. From top to bottom, $\tautr = 10, 5,
2.5,$ and $1.0$.}
\label{fig:sequence-of-spectra}
\end{figure*}

The model contains three regions: (1) A cold accretion disk, assumed here
to have a temperature $kT_{\rm bb} = 150$ keV, (2) the transition layer,
situated directly above the cold disk, and (3) the corona, situated
directly above the transition layer. Plane parallel geometry is assumed.

We use the slab-geometry ADC model of Dove, Wilms, \& Begelman (1997),
which uses a non-linear Monte Carlo (NLMC) routine to solve the radiation
transfer problem of the system.  The free parameters of the model are the
seed optical depth $\taue$, (the optical depth of the corona excluding the
contribution from electron-positron pairs), the blackbody temperature of
the accretion disk and its compactness parameter, $\lbb$, and the heating
rate (i.e., the compactness parameter), $\lc$, of the ADC.  The temperature
structure of the corona is determined numerically by balancing Compton
cooling with heating, where the heating rate is assumed to be uniformly
distributed.  The $e^-e^+$-pair opacity is given by balancing photon-photon
pair production with annihilation.  Reprocessing of coronal radiation in
the cold accretion disc is also treated numerically. For a more thorough
discussion of the NLMC routine, see Dove, Wilms, \& Begelman (1997). The
transition layer is treated identically to the corona, accept here the
heating rate is set to zero. Therefore, the transition layer, numerically
modeled using 8 shells, each with equal optical depth ${\rm d}\tau =
\tautr/8$, will obtain the Comptonization temperature due to the
radiation field from both the corona and the accretion disk.
  
\subsection{Maximum Coronal Temperature}

As discussed by Dove, Wilms, \& Begelman (1997), for a given total
optical depth, there exists a maximum coronal temperature, above which
no self-consistent solution exists. Raising the compactness parameter
of the corona to a value higher than that corresponding to the maximum
temperature gives rise to a higher optical depth (due to pair
production), causing more reprocessing, subsequent Compton cooling,
and ultimately a {\em lower} coronal temperature. Here, as much as
80-90\% of the incident X-ray flux is re-radiated at the disk
temperature, i.e., the X-ray integrated albedo is 0.1-0.2.  A
completely ionized transition layer will increase the albedo of the
cold disk.  Accordingly, the amount of energy in the reprocessed
blackbody (or Comptonized blackbody) should become smaller with
increasing Thomson optical depth of the transition layer, and this
Compton cooling via reprocessed radiation should be less efficient and
therefore higher maximum coronal temperatures (as compared to models
with no transition layer) should be allowed.

\begin{figure*}
\centerline{\psfig{file=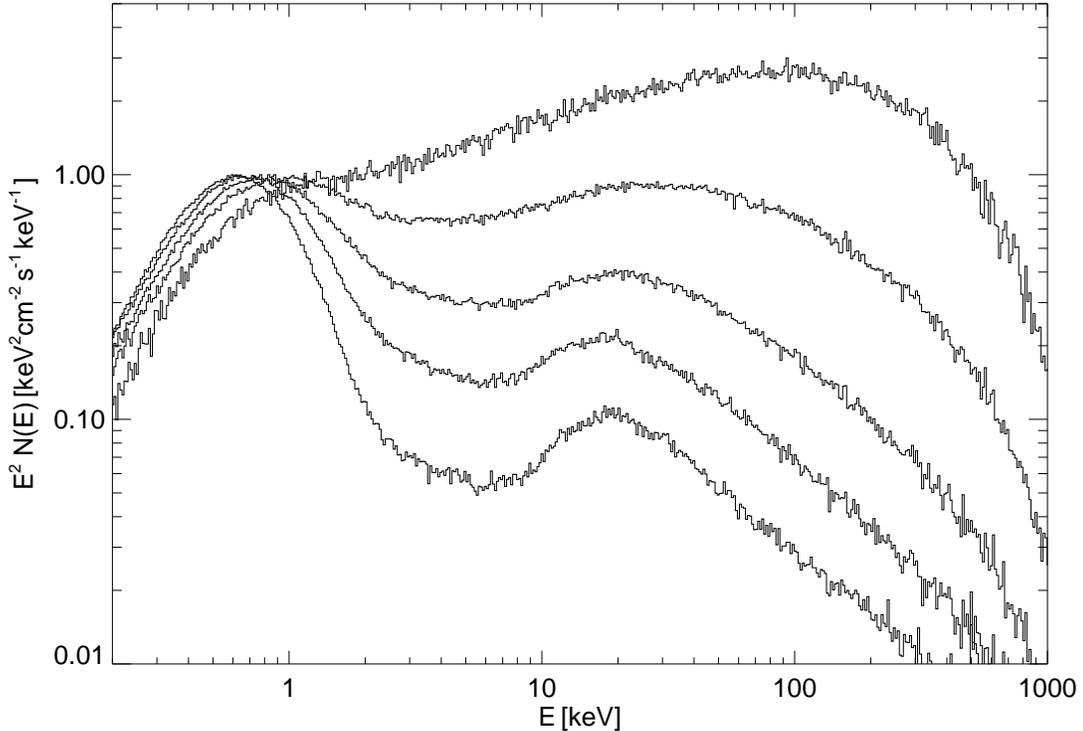,width=.8\textwidth}}
\caption{Internal spectrum for several shells within the transition disk
  and corona. From bottom to top, $\tautr(z) = 1.25, 2.5, 3.75,$ and $
  5.0$, where $\tautr(z)$ is the Thomson optical depth from the
  cold-disk/transition layer interface ($z=0$) to a height $z$. The
  uppermost spectrum is the internal spectrum within the corona. For this
  model, $\tautr = 5.0$ and $\tauc = 0.3$.}
\label{fig:spec-vs-tau}
\end{figure*}

In Figure \ref{fig:Tmax-vs-tautrans}, we show how the average coronal
temperature varies with the optical depth of the transition layer,
$\tautr$. In Figure \ref{fig:sequence-of-spectra}, we also show the
resulting broad band spectra for the model parameters tested. For all
models, we chose $\lbb = 0.01$ in order to be consistent with the
definition of the disk compactness parameter, i.e., $\lbb\equiv
\fdisk\sigma_T H/ ( m_e c^3)$, according to which
\begin{equation}
\lbb = 0.03 \left({\dm\over 0.05}\right)^{6/5}
\left(1-f\right)^{11/10} \left(\alpha M_1\right)^{-1/10}\; .
\label{lbb}
\end{equation}
Further, other parameter values are $\lc = 2$, $kT_{bb} = 150$ keV,
and $\tau_c = 0.3$. These parameters correspond to the model producing
the maximum coronal temperature. In contrast to models in which
$\tau_{\rm tr} = 0$, the coronal temperature for a given value of
$\tau_c$ is not simply a function of $\lc/\lbb$. To see this, consider
the case where $\tautr \gg 1$. Here, the albedo of the disk is
essentially unity, and therefore all of the soft photons emitted will
be from the intrinsic flux of the disk (no reprocessing). Therefore,
setting $\lbb \ll 1$ yields the maximum coronal temperatures possible.
Note that the maximum temperature levels out as $\tautr \rightarrow
\infty$. Although, in this limit, there is no reprocessing of hard
X-rays in the cold disk, there is still ``reprocessing'' in the
transition layer.  As $\tautr$ increases, more coronal radiation is
down-scattered to the Compton temperature of the transition layer,
which is $kT_{tr} \sim 1 - 4$ keV. Even at these temperatures, Compton
cooling of this ``reprocessed'' radiation in the corona is very
efficient.

It is interesting to note that, only for $\tau_{\rm tr} \simgt 10$,
the coronal temperature is high enough such that the corresponding
spectrum of escaping radiation is hard enough to describe the
observations of Cyg~X-1.  (The canonical value of the photon power law
of Cyg~X-1 is $\Gamma = 1.7$; for $\tau_c = 0.3$, this power law
corresponds to $kT_c \sim 150$ keV). It is probably unphysical,
however, to assume the transition layer is completely ionized for such
large optical depths. In fact, the numerical model for $\tau_{\rm tr}
= 10$ predicts a temperature of $kT_{\rm tr} \sim 500$ eV near the
bottom of the layer. Furthermore, as is seen in Figure
\ref{fig:spec-vs-tau}, the radiation spectra do become substantially
softer deeper in the transition layer, and, using our simple
experimentation with softer ionizing spectra, shown in Figure
\ref{fig:ssoft}, we can expect that the gas will become stable and may
be at a temperature lower than the Compton temperature for $\tau_{\rm
tr}$ as small as $2-3$. Therefore, even with the advent of transition
layers, it still appears unlikely that a global slab geometry ADC
model can have self-consistent temperatures high enough to reproduce
the observed hard spectra of Cyg X-1 and other similar BHCs. This does
not rule out ADC models in which the corona is patchy, e.g.,
containing several localized active regions such as magnetic flares.
The patchy geometry leads to higher coronal temperatures due to less
reprocessed soft flux re-entering active regions (Poutanen \& Svensson
1996), so that lower $\tau_{\rm tr}$ may be sufficient to explain the
Cyg X-1 spectrum.

\subsection{Reprocessing Features in the Spectrum of Escaping Radiation}

As shown in Figure \ref{fig:sequence-of-spectra}, the reprocessing
features in the spectrum of escaping radiation depend sensitively on
the optical depth of the transition layer. In Figure
\ref{fig:spec-vs-tau}, we show how the internal spectrum (averaged
over all angles) varies throughout the transition layer and corona.
Note that most of the hard X-rays do not penetrate through the
transition layer, and that the spectrum gets softer as it approaches
the cold disk. On the other hand, the thermal radiation emitted by the
disk and the 10-50 keV Compton reflection hump are broadened by
Compton scatterings as the disk radiation diffuses upward through the
transition layer and the corona. The Iron K$\alpha$ line, small to
start with due to the small amount of reprocessing of coronal
radiation, is completely smeared out by the time the radiation escapes
the system.  No line photons are created in the transition layer
itself, because we found that the Compton equilibrium state typically
resides at the ionization parameter $\xi \simgt 10^4$, whereas no
fluorescent iron line emission is produced for $\xi \simgt 5\times
10^3$ (Matt, Fabian \& Ross 1993, 1996).

Therefore, in agreement with B\"ottcher, Liang, and Smith (1998), we
conclude that a highly ionized transition layer can reduce the
reprocessing features in the spectrum of escaping radiation emanating
from slab geometry ADC models. Previous work has used the lack of
observed reprocessing features to argue against the applicability of
slab geometry ADC models, and argued for a disk/corona configuration
where the disk has a small solid angle relative to the coronal region,
such as a ``sphere$+$disk model (Dove et al. 1998, Gierlinski et
al. 1997, Poutanen, Krolik \& Ryde 1997). A slab geometry ADC model
containing a transition layer, however, is another possibility.
Although, as discussed above, these models could not obtain high
enough temperatures to be able to predict spectra as hard as the
observed spectra of GBHCs unless $\tautr \gta 10$, a model with a
patchy corona and underlying transition layers appears likely to be
able to predict hard enough spectra with a relatively small amount of
reprocessing features. In a forthcoming paper, we plan to investigate
such a model.

\section{Discussion}\label{sec:discussion}

We have shown that the transition layer in the vicinity of transient
flares for ADC models of GBHCs must be highly ionized and very
hot. Specifically, due to a thermal instability, the only stable
temperature of the transition layer is the local Compton temperature
($k T\sim$ few keV).  In fact, even for global ADC models of GBHCs,
such a transition layer is found to be likely. However, we have not
accurately modeled the full radiative transfer problem through the
transition layer at this time, and therefore could not determine its
vertical optical depth, which was treated as a free parameter.

Due to the transition layer, a larger fraction of incident X-rays are
Compton reflected back into the corona without being reprocessed by the
cold disk. In addition, for $\tau_{\rm trans}\gg 1$, the predicted
reprocessing features as well as the thermal excess should be substantially
smaller (as is found in this paper for the global ADC model) than that of
previous ADC models in which the transition layer was not considered.  This
reduction of the reprocessing features is crucial for the model being
consistent with the observations of GBHCs (e.g., Gierlinski et al.  1997,
Dove et al. 1998).

This reduction in reprocessing yields a lower Compton cooling rate within
the corona, and higher coronal temperatures than previous ADC models are
allowed.  For global ADC models with $\tauc \sim 0.3$, we find that the
coronal temperature can be as high as $\sim 150$ keV {\em if} the optical
depth of the transition layer is $\tautr \gta 10$.  However, a completely
ionized transition layer with $\tautr \gta 10$ is most likely physically
inconsistent since the bottom layer was found to be too cold to
realistically stay highly ionized. Therefore, global slab-geometry ADC
models are still problematic in explaining the observations of GBHCS.
Nevertheless, these global ADC results are very encouraging, and a slab
geometry ADC model containing a patchy corona (e.g., individual ARs) with
underlying transition layers should be rigorously studied. Here, due to the
lower amount of reprocessed radiation within the ARs as compared to the
global model, models with coronae hot enough to reproduce the observed
spectra of GBHCs may be allowed for more reasonable transition layer
optical depths.

We have considered only radiative cooling mechanisms for the
transition layer in this paper.  It is possible that a wind is induced
by the X-ray heating.  However, the maximum gas temperature obtained
due to the X-ray heating is the Compton temperature ($\lesssim 10^8$
K). Therefore, as shown by Begelman, McKee \& Shields (1983), a large
scale outflow cannot occur for $R\lesssim 10^4 R_g$. On the other
hand, a local uprising of the gas is still possible. The maximum
energy flux due to this process is $F_{\rm ev} \sim P c_s$, where
$c_s$ is the sound speed in the transition region. Since $P\lesssim
\fx/c$, and $c_s \lesssim 3\times 10^{-3}\, c$, we have ${F_{\rm
ev}/\fx} \lesssim 3 \times 10^{-3}$. Therefore, a wind {\em or any
other mechanical process} cannot cool the gas efficiently, and thus it
is justifiable to consider cooling via emission of radiation only.

In a forthcoming paper, we will discuss the implications of the
instability for the case of AGN. As shown in Figures \ref{fig:scurve}
and \ref{fig:ssoft}, there is a stable ionization equilibrium state
below $T\sim 3\times 10^5$ Kelvin. Furthermore, due to increasing
opacity to the soft disk radiation, we find that equilibria below
$T\sim 10^5$ K are not possible.  We found that the existence of this
stable region, and the unstable region above $3\times 10^5$ K, is very
much independent of the details of the incident spectra (indeed,
Figures \ref{fig:scurve} and \ref{fig:ssoft} were not even intended
for the AGN case). Thus, it is possible that the differences in the
intrinsic X-ray spectra, the ionization state of the reflector and the
strength of the iron line in AGNs and GBHCS can be explained by the
different end points of the thermal ionization instability described
here.

\section{Acknowledgments}

SN acknowledges the contributions, guidance and financial support 
through NASA grant NAG 5-3075 by Prof. Fulvio Melia. JD acknowledges the
useful discussions with P. Maloney, J. Wilms, and M. Nowak.

{}


\begin{thebibliography}{}

\bibitem[]{} \apj{Begelman, M.C., McKee, C.F., \& Shields, G.A.
1983}{271}{70}

\bibitem[]{} \refbook{B\"ottcher, M., Liang, E. P., \& Smith, I. A.,
  1998, submitted to A\&A.}

\bibitem[]{}\refbook{Brandt, W.N., \& Boller, Th. 1998, preprint,
astro-ph/9808037}
  
\bibitem[Dove, Wilms, \& Begelman 1997]{dove97a} \apj{Dove, J. B., Wilms,
    J., \& Begelman, M. C., 1997}{487}{747}

\bibitem[Dove et al. 1998]{dove98}\mnras{Dove, J. B., Wilms, J., Nowak,
M. A., Vaughan, M. A., 1998}{298}{729}

\bibitem[]{}\apj{Field, G.B. 1965}{142}{531}


\bibitem[Galeev, Rosner \& Vaiana 1979] {galeev} \apj{
Galeev, A. A., Rosner, R., \& Vaiana, G. S., 1979 }{ 229}{ 318}

\bibitem[]{} \mnras{Gierlinski, M. et al. 1997}{288}{958}

\bibitem[Haardt \& Maraschi 1991] {hm91} \apj{Haardt, F., \& Maraschi, L.
1991} {380}{ L51}
\bibitem[Haardt \& Maraschi 1993] {hm93} \apj{Haardt, F., \& Maraschi, L.
1993}{413}{ 507}
\bibitem[Haardt et al. 1994] {haardt94} \apj{Haardt, F., Maraschi, L.,
\& Ghisellini, G. 1994}{432}{ L95}

\bibitem[Kallman \& McCray 1982]{kallman82} \apjsup{Kallman, T. R., \& McCray,
    R. 1982}{50}{263}

\bibitem[]{} \refbook{Kallman, T.R., \& Krolik, J.H., 1986, NASA/GSFC
Laboratory for High Energy Astrophysics special report.}

\bibitem[]{}\apj{Krolik, J.H., McKee, C.F., \& Tarter, C.B. 1981}
{249}{422}

\bibitem[]{} \apj{Maciolek-Niedzwiecki, A., Krolik, J.H., \&
Zdziarski, A.A. 1997}{483}{111}

\bibitem[]{}\mnras{Magdziarz, P. \& Zdziarski, A.A. 1995}{273}{837}

\mnras{Matt, G., Fabian, A.C., \& Ross, R. R. 1993}{262}{179}

\mnras{Matt, G., Fabian, A.C., \& Ross, R. R. 1996}{278}{1111}


\bibitem[Nayakshin \& Melia 1997]{} \apjlett{Nayakshin, S. \& 
Melia, F. 1997}{484}{L103}

\bibitem[]{}\refbook{Nayakshin S. 1998, Ph.D. Thesis, The University
of Arizona. Also available on astro-ph archive.}

\bibitem[] {np94} \mnras{Nandra, K. \& Pounds, K. 1994}{268}{405}

\bibitem[] {p90} \nature{Pounds, K.A. et al. 1990}{344}{132}


\bibitem[]{} \mnras{Poutanen, J., Nagendra, K.N., \& Svensson, R. 1996}
{283}{892}

\bibitem[]{} \apj{Poutanen, J. \& Svensson, R. 1996}{470}{249}

\bibitem[]{} Poutanen, J, Krolik, J.H, \& Ryde, F. 1997, in the Proceedings of
the 4th Compton Symposium, astro-ph/9707244

\bibitem[]{} \refpaper{Poutanen, J., Svensson, R., \& Stern,
B. 1997}{in the Proceedings of the 2nd INTEGRAL Workshop ``The
Transparent Universe''}{ESA SP-382}{}

\bibitem[]{} \mnras{Ross, R.R., Fabian, A.C., \& Brandt,
W.N. 1996}{278}{1082}

\bibitem[Sincell and Krolik 1997] {sk97} \apj{Sincell, M.W. \&
Krolik, J.H. 1997}{476}{605S}

\bibitem[Stern et. al. 1995]{stern95b}\apj{Stern, B., Poutanen, J.,
    Svensson, R., Sikora, M., \& Begelman, M. C. 1995}{449}{L13}

\bibitem[Svensson \& Zdziarski 1994] {sz94}\apj{Svensson, R. \&
Zdziarski, A. 1994}{ 436}{ 599}

\bibitem[Svensson 1996] {sv96} \aasup{Svensson, R. 1996}{120}{475}

\bibitem[]{zd96} \aasup{Zdziarski, A.A. et al. 1996}{120}{553}

\bibitem[]{} \apj{\'Zycki, P.T. et. al. 1994}{437}{597}


\end{thebibliography}
\end{document}